\documentclass[aps,twocolumn,showpacs,amsmath,amssymb]{revtex4}
\usepackage{amsthm}
\usepackage{latexsym}
\usepackage{amsfonts}
\usepackage{bbm,dsfont}
\usepackage{graphicx}

\newtheorem{definition}{Definition}

\newtheorem{proposition}{Proposition}

\newcommand{\h}[1]{\mathcal{#1}}
\newcommand{\R}{\mathbb{R}}
\newcommand{\C}{\mathbb{C}}
\newcommand{\N}{\mathbb{N}}
\newcommand{\hil}{\mathcal{H}}
\newcommand{\br}{\mathcal{B}(\R)}

\newcommand{\lh}{\mathcal{L(H)}}
\newcommand{\tc}{\mathcal{T(H)}}

\newcommand{\sfp}{\mathsf{P}}

\newcommand{\p}{\mathsf{p}}
\newcommand{\E}{\mathsf{E}}
\newcommand{\F}{\mathsf{F}}
\newcommand{\sfz}{\mathsf{Z}}

\newcommand{\M}{\mathsf{M}}
\newcommand{\tr}[1]{\mathrm{tr}\left[ {#1} \right]}

\begin{document}

\title{Weak vs. approximate values in quantum state determination}

\author{Erkka Haapasalo}
\email{erkka.haapasalo@utu.fi}
\affiliation{Turku Centre for Quantum Physics, Department of Physics and Astronomy, University of Turku, FI-20014 Turku, Finland}

\author{Pekka Lahti}
\email{pekka.lahti@utu.fi}
\affiliation{Turku Centre for Quantum Physics, Department of Physics and Astronomy, University of Turku, FI-20014 Turku, Finland}

\author{Jussi Schultz}
\email{jussi.schultz@utu.fi}
\affiliation{Turku Centre for Quantum Physics, Department of Physics and Astronomy, University of Turku, FI-20014 Turku, Finland}

\date{\today}
\begin{abstract}
We generalize the concept of a weak value of a quantum observable to cover arbitrary real  positive operator measures. We show that the definition is operationally meaningful in the sense that it can be understood within the quantum theory of sequential measurements. We then present a detailed analysis of the recent experiment of Lundeen {\em et al.} \cite{Lundeen2011} concerning the reconstruction of the state of a photon using weak measurements. We compare their method with the reconstruction method through informationally complete phase space measurements and show that it lacks the generality of the phase space method. In particular, a completely unknown state can never be reconstructed using the method of weak measurements. 
\end{abstract}

\maketitle

\section{Introduction}
The weak value of a quantum observable, as defined by Aharonov {\em et al.} \cite{Aharonov1988}, has always remained a somewhat controversial concept despite the vast amount of attention it has received over the years (see, e.g., \cite{Aharonov2005} and references therein). The purpose of this paper is to clarify this topic and our goal is 
two-fold. On one hand, we wish to show that the weak values can be exhaustively explained within the quantum theory of sequential measurements. 
For that end we briefly recall the quantum theory of measurement (Section \ref{measurements}) and generalize the standard measurement model such that it can be applied to arbitrary real observables 
(as positive operator measures) (Section \ref{standard_model}). We then use the theory of sequential measurements to define a weak value of an observable (Section \ref{weak_values}) and compare it  with
its approximative values obtained from the  relevant 
standard measurement.  On the other hand, in the view of the recent paper of Lundeen {\em et al.} \cite{Lundeen2011} we wish
to compare the role of weak and approximate measurements in quantum state determination (Section \ref{reconstruction}). We analyze the method of Lundeen {\em et al.} and compare it with the known method of approximate sequential measurements.  

Throughout the paper we denote by $\hil$ the Hilbert space associated with a physical system, and by $\lh $ and $\tc$ the sets of bounded and trace class operators acting on $\hil$. 
The concepts of states, observables, and the statistical duality they define form the rudimentary frame of the  description of the system: a state $\rho$ is (represented as) a positive ($\rho \geq 0$) trace one ($\tr{\rho}=1$) operator, an observable is (represented as) 
a normalized positive operator measure (POM) $\E:\h A\rightarrow \lh$ (defined on a $\sigma$-algebra $\h A$ of subsets of a set $\Omega$ of the values of the observable), the probability measure $\h A\ni X\mapsto\p^\E_\rho(X) =\textrm{tr}[\rho \E(X)]$ giving the measurement outcome statistics for the observable $\E$ in the state $\rho$. Usually the value space $(\Omega,\h A)$ of an observable is just the real Borel space $(\R,\h B(\R))$, which we use subsequently. For a pure state ($\rho=\rho^2$) we use also a unit vector (vector state) representation $\varphi\in\hil$,
$\rho\varphi=\varphi$.

\section{Quantum measurements}\label{measurements}

Quantum theory of measurement operates on a hierarchy of three levels of description reflecting the options of  restricting one's attention
to the outcome probabilities at the level of the measured system, or taking into account system's conditional state changes, or adopting the most comprehensive level of modelling the interaction and information transfer between the system and the probe.
We shall briefly describe these three levels since all of them play a role in our analysis, for a more comprehensive survey, see e.g. \cite{QTM}.

The crudest statistical level is the one described in Introduction. On the next level, one also describes the conditional state changes of the system due to a measurement, conditioned with respect to the pointer values. These changes are  most conveniently captured in the concept of an {\em instrument} \cite{Davies1970,Davies1976}, that is, an operation valued measure $\h I :\h B(\R)\rightarrow \h L (\tc)$. This means that each $X\in \h B(\R)$ determines a positive contractive linear map $\h I (X) :\tc \rightarrow \tc$ and $\h I (X)(\rho)$ is the unnormalized conditional output state.  Each instrument determines uniquely the associated observable via the formula $\textrm{tr}[\rho \E(X)] = \textrm{tr}[\h I(X) (\rho)]$, or equivalently via the dual instrument as $\E (X) = \h I(X)^* (I)$.

The most detailed descriptions of measurements are given by the  {\em measurement schemes} where the coupling of the physical system and the probe (aka measuring apparatus) is considered. More precisely, a measurement scheme is a 4-tuple $\h M =\langle  \h K  , \sigma , \Phi ,  \sfz\rangle$, where $\h K$ is the Hilbert space of the probe, $\sigma$ its initial state, $\Phi :\h T (\hil \otimes \h K) \rightarrow \h T (\hil \otimes \h K)$ a state transformation (i.e. a trace-preserving operation) modelling the interaction between the system and the probe, and $\sfz:\h B(\R)\rightarrow\h L (\h K)$  the pointer observable (POM). Each measurement scheme determines a unique instrument via
\begin{equation}\label{instru}
\h I(X) (\rho)  = \textrm{tr}_{\h K} [\Phi (\rho \otimes \sigma) I \otimes \sfz (X)]
\end{equation}
where $\textrm{tr}_{\h K}[\cdot]$ denotes partial trace over $\h K$. As a consequence, a measurement scheme also determines uniquely the measured observable $\E$, and one has
\begin{equation}\label{prc}
 \textrm{tr}[\rho \E(X)]= \textrm{tr}[\h I(X) (\rho)]= \textrm{tr} [\Phi (\rho \otimes \sigma) I \otimes \sfz (X)].
\end{equation}
In most cases $\Phi$ can be given by a unitary operator $U$  on $\h H\otimes\h K$ so that $\Phi(\rho \otimes \sigma)= U(\rho \otimes \sigma)U^*$. It is often convenient to allow the pointer observable $\sfz$ to have different values than the measured observable $\E$ has; in such a case one needs to introduce a (measurable)  pointer function $f:\R\to\R$ and adjust the   conditions \eqref{instru} and \eqref{prc} accordingly, that is, replacing the  set $X$ with $f^{-1}(X)$ for the pointer observable $\sfz$. The fundamental result of the quantum measurement theory implies that each observable $\E$ admits a measurement scheme in the sense of Eq. \eqref{prc} \cite{Ozawa1984}.

We wish to emphasize that the correspondences between measurement schemes and instruments, and instruments and observables are many-to-one, reflecting the obvious fact that a given observable may be measured in various ways, as well as that a given instrument may be realized by various measurement schemes.

Now suppose that one  wants to measure a pair of observables 
$\E_1$ and $\E_2$ by performing their measurements sequentially, for instance, first $\E_1$ and then $\E_2$. 
On the level of instruments this leads to the sequentially composed instrument $\h I_{12}$ defined by
$\h I_{12}(X\times Y) = \h I_2 (Y) \circ \h I_1 (X)$ \cite{Davies1970,Busch1990}. 
On the statistical level such a sequential measurement defines
the (sequential joint) observable $\M$ given by $\M ( X\times Y) = \h I_1(X)^*(\h I_2 (Y)^*(I))$. In particular, the Cartesian margins of $\M$ are 
\begin{align}
\M_1 (X) &=\M(X\times \R) = \E_1 (X),\nonumber\\
\M_2 (Y) &= \M(\R\times Y) = \h I_{1} (\R)^* ( \E_2 (Y))\nonumber
\end{align}
which shows the characteristic quantum feature that the first measurement typically disturbs the subsequent one: the first margin is the observable measured first, the second margin is a smeared (or disturbed) version of the observable measured second, smearing depending on the first measurement. It is to be emphasized that the structure of a sequential joint observable $\M$ does not depend on any details of the second measurement.

\section{Generalized standard model}\label{standard_model}

As an example (and for later use) we present a generalization of the standard model of a measurement. This model goes back to von Neumann \cite{vN}, for a survey, see \cite{Busch1996}.

Let $\E:\h B(\R)\rightarrow \lh$ be an observable to be measured. By a special version \cite{Ozawa1984} of the Naimark dilation theorem there exists a Hilbert space $\hil_0$, a unit vector $\psi\in\hil_0$ and a spectral measure $ \sfp^A:\h B(\R)\rightarrow \h L(\hil\otimes \h K)$, with the corresponding selfadjoint 
operator $A$, such that 
$$
\E (X) = V_\psi^*\, \sfp^A (X)\, V_\psi
$$
where $V_\psi:\hil \rightarrow\hil\otimes \hil_0$ is the embedding $V_\psi (\varphi ) =\varphi\otimes \psi$, $\varphi\in\hil$. 

Let $\h K =L^2 ( \R)$ be the Hilbert space of the probe. For each  $\lambda >0$ define the state transformation $\Phi^\lambda: \h T(\hil \otimes \h K)\rightarrow   \h T(\hil \otimes \h K)$ via
$$
\Phi^\lambda (\rho \otimes \sigma) =\textrm{tr}_{\hil_0} [e^{-i\lambda A \otimes P} \rho\otimes \vert\psi\rangle\langle \psi\vert  \otimes \sigma e^{i\lambda A \otimes P} ]
$$
where $P$ is the momentum operator in $\h K$. Since $P$ generates spatial translations, it is natural to choose as the pointer observable the position  of the probe, represented by the spectral measure $\sfp^Q:\h B(\R)\rightarrow \h L(\h K)$ of the position operator $Q$. Due to the coupling constant $\lambda$ it is now convenient to choose a pointer function $f^\lambda(x) = \lambda^{-1} x$. The 5-tuple $\h M^\lambda = \langle \h K,  \sigma,  \Phi^\lambda,  \sfp^Q, f^\lambda \rangle$ constitutes a measurement scheme with the intention to measure the system observable $\E$. We call this a {\em generalized standard model for $\E$}.

The instrument  as well as the observable actually measured are now easily computed from (\ref{instru}) and (\ref{prc}) with the adoption of the pointer function $f^\lambda$. Since we consider their $\lambda$-dependence we explicitly parametrize the instrument as well as the observable by that only. For notational simplicity, we assume that the initial probe state $\sigma$ is a pure state given by a unit vector $\phi$. One obtains the associated instrument and its dual
\begin{eqnarray*}
\h I^\lambda (X) (\rho) &=& \int_X \textrm{tr}_{\hil_0} [K_x V_\psi \rho V_\psi^* K_x^*]\, dx, \quad \rho\in\tc,\\
\h I^\lambda (X)^* (B) &=& \int_X V_\psi^* K_x^* (B\otimes I) K_x V_\psi\, dx, \quad B\in\lh,
\end{eqnarray*}
where
$$
K_x = \sqrt{\lambda} \phi (-\lambda (A-x)) =\int\sqrt{\lambda} \phi (-\lambda (y-x))\, d\sfp^A(y),
$$
for all $x\in\R$. The observable actually measured by this scheme is a smeared version $\mu^\lambda *\E$ of the desired one, where the convolution is defined as
$$
\mu^\lambda *\E (X) = \int \mu^\lambda (X-x) \, d\E(x), \quad X\in\h B(\R),
$$
and the convolving probability measure is defined through $\mu^\lambda(X)=\langle\phi\vert\sfp^Q(\lambda X)\phi\rangle$.

We wish to emphasize that although the generalized standard measurement model as well as the associated instrument depend on the used Naimark dilation of $\E$ (aka ancilla) the actually measured observable $\mu^\lambda*\E$ is independent of the used dilation.
In particular, this result remains tacit if $\E$ is a spectral measure (in which case no ancilla is needed).

The (generalized) standard model for an observable $\E$ constitutes its {\em approximate measurement} in the sense that the actually measured observable $\mu^\lambda*\E$ is a convolution of $\E$ with the probability measure $\mu^\lambda$. The  approximation depends on $\mu^\lambda$ and the degree of approximation can be quantified in many different ways, for instance, by the standard deviation of $\mu^\lambda$.

The set of possible values, that is, the set of possible measurement outcomes of the observable $\mu^\lambda*\E$, namely the support of the POM, ${\rm supp}(\mu^\lambda*\E)$, is typically much bigger than the set of possible values of $\E$, ${\rm supp}(\E)$. For instance, if $\E$ is a two-valued spectral measure, ${\rm supp}(\mu^\lambda*\E)$ can be anything from this two-point set till the whole of $\R$, depending on the support of $\mu^\lambda$.  We call the numbers in ${\rm supp}(\mu^\lambda*\E)$ the {\em approximate values} of $\E$ obtained from the (generalized) standard model $\h M^\lambda$. We also note that if one chooses the `smearing' measure $\mu^\lambda$ such that its expectation (or average) value is zero then the actually measured observable $\mu^\lambda*\E$ and the observable $\E$ are statistically indistinguishable in the level of first moments (expectation values). However, the actual observables (POMs) coincide if and only if $\mu^\lambda$ is a point measure concentrated at the origin, a condition which is clearly impossible for our specific measurement scheme $\h M^\lambda$.

\section{Generalized weak values}\label{weak_values}

The sequential measurement scheme can be used to define what Aharonov {\em et al.} called the weak value of an observable \cite{Aharonov1988}. The intuitive idea behind the weak value is that by letting the strength of the interaction between the object and probe become sufficiently weak, the disturbance caused by the first measurement on the system becomes negligible. The price to be paid is that the first measurement becomes very poor. In other words, the observable becomes more and more smeared. However, by a clever choice of the probe state it is possible to control the measurement so that the average value of the first measurement remains the same.

The original line of reasoning can be generalized to cover arbitrary pairs of observables. In order to make it rigorous, we need some technical details. Let $\E:\br\rightarrow \lh$ be an observable. For all $\psi,\varphi\in\hil$ denote by $\E_{\psi,\varphi}$ the complex measure $X\mapsto \langle\psi \vert \E(X)\varphi\rangle$. Let $\h D(x,\E)\subset\hil$ denote the subspace of those $\varphi$ for which the identity map $x\mapsto x$ is $\E_{\psi,\varphi}$-integrable for all $\psi\in\hil$. The first moment operator of $\E$ is then the  linear operator $\E[1]:\h D(x,\E)\rightarrow \hil$ defined as 
$$
\langle\psi \vert\E[1]\varphi\rangle  =\int x\, d\E_{\psi,\varphi}(x),\qquad \varphi\in\h D(x,\E), \psi\in\hil.
$$
The domain of $\E[1]$ contains as a subspace the square-integrability domain $\widetilde{\h D}(x,\E)$ consisting of those $\varphi\in\hil$ for which the function $x\mapsto x^2$ is integrable with respect to the positive measure $\E_{\varphi,\varphi}$.
In general these two domains are different but it may happen that they coincide. This is the case for example for spectral measures, i.e., sharp observables in which case we simply have $\sfp^A [1] =A$.

\begin{definition}\label{Def}
Let $\E,\F:\br\rightarrow \lh$ be observables, $\varphi\in \h D(x,\E)$, $\Vert \varphi\Vert=1$ and let $Y\in\br$ be such that $\F(Y)\varphi\neq 0$.
The weak value of $\E$ in a vector state $\varphi$ conditioned by $\F(Y)$ is
\begin{equation}\label{new_weak_value}
\E_w (\varphi,\F(Y)) = \frac{\big\langle \varphi\big\vert \F(Y) \E[1]\varphi \big\rangle}{\big\langle
\varphi\big\vert \F(Y)\varphi\big\rangle}
\end{equation}
\end{definition}

Notice that if $\E$ is of the form $\mu*\sfp^A$ for a probability measure $\mu$, with $\mu[1]=\int x\,d\mu(x)=0$, so that $\E[1]=A$, and  $\F$ is a discrete observable with $\F(Y) = \vert \eta \rangle\langle \eta\vert$ for some unit vector $\eta\in\hil$, we obtain the familiar expression
$$
\E_w (\varphi,\F(Y)) =\frac{\big\langle \varphi\big\vert \F(Y) \E[1]\varphi \big\rangle}{\big\langle\varphi\big\vert \F(Y)\varphi\big\rangle} =  \frac{\big\langle \eta\big\vert A\varphi \big\rangle}{\big\langle\eta\big\vert \varphi\big\rangle}.
$$

We will now proceed to showing that the general definition is operationally meaningful. Indeed, we will show that the real and imaginary parts can be obtained as conditional averages in two different sequential measurement schemes in the limit of zero interaction strength.

Consider the generalized standard measurement scheme $\h M^\lambda$ and suppose that after this measurement, realizing the observable $\mu^\lambda *\E$, we perform (an exact) measurement of $\F$ thus obtaining the (sequential joint) observable $\M^\lambda:\h B(\R^2)\rightarrow \lh$. If we then postselect only the values $(x,y)$ with $y\in Y$ (for a fixed $Y$ for which $\M^\lambda_2 (Y)\varphi \neq 0$) and normalize the probabilities we end up with a conditional probability measure
$$
X\mapsto \frac{\big\langle \varphi \big\vert \h I^\lambda (X)^* (\F (Y)) \varphi \big\rangle}{\big\langle \varphi \big\vert \h I^\lambda (\R)^* (\F (Y)) \varphi \big\rangle}=\frac{\big\langle \varphi \big\vert\M^\lambda(X\times Y) \varphi \big\rangle}{\big\langle \varphi \big\vert \M_2^\lambda (Y) \varphi \big\rangle} 
$$
We now claim that the real part of the weak value is obtained by taking the average of the above measure and then taking the limit of zero interaction strength, that is,
\begin{equation}\label{limit}
\textrm{Re}\, \big( \E_w (\varphi,\F(Y))\big) =\lim_{\lambda\rightarrow 0} \int x\,   \frac{\big\langle \varphi \big\vert \h I^\lambda (dx)^* (\F (Y)) \varphi \big\rangle}{\big\langle \varphi \big\vert \h I^\lambda (\R)^* (\F (Y)) \varphi \big\rangle}    .
\end{equation}

First notice that $\lim_{\lambda \rightarrow 0}  \big\langle \varphi \big\vert \h I^\lambda (\R)^* (\F (Y)) \varphi \big\rangle = \langle \varphi \vert \F(Y) \varphi\rangle$ so it is sufficient to consider only the numerator for which we define $\Lambda^\lambda (X) = \big\langle \varphi \big\vert \h I^\lambda (X)^* (\F (Y)) \varphi \big\rangle$. Now by assuming that $\phi\in \h D(Q)$ and $\varphi\in \widetilde{\h D}(x,\E)$ (the latter being equivalent to $\varphi\otimes\psi\in\h D(A)=\h D(x,\sfp^A)$), and by using the translation covariance of position we can show that $e^{-i\lambda A\otimes P} \varphi \otimes \psi \otimes \phi \in\h D(x,I\otimes I\otimes \sfp^Q)$ 
for all $\lambda \geq 0$. It follows that the required average value is given by 
$$
\Lambda^\lambda [1] =\frac{1}{\lambda} \big\langle   e^{-i\lambda A\otimes P} \varphi\otimes \psi\otimes\phi \big\vert  I\otimes I\otimes Q  e^{-i\lambda A\otimes P} \varphi\otimes \psi\otimes\phi \big\rangle
$$
If we then make the assumptions that $\langle \phi \vert  Q\phi \rangle =0 $ and $\phi \in \h D(QP) \cap \h D(PQ)$  we can calculate the limit $\lambda \rightarrow 0$. Using again the translation covariance of position, we can actually calculate the limit in a rigorous manner and obtain
\begin{align}
 \lim_{\lambda \rightarrow 0} \Lambda^\lambda [1] =&\,  i   \, \langle \E[1]\varphi  \vert \F(Y)\varphi \rangle \langle \phi \vert PQ\phi \rangle  \nonumber\\
& -i\,  \langle \F(Y)\varphi  \vert \E[1]\varphi  \rangle \langle  \phi \vert QP\phi \rangle .
\end{align}
Here we have used the fact that the operator identity $V_\psi^* A V_\psi =\E[1]$ holds on the square-integrability domain $\widetilde {\h D}(x,\E)$ \cite{Lahti1999}.

We immediately see that a sufficient condition for the above limit to be proportional to the real part of the weak value is that $\langle \phi \vert \{Q, P\}\phi\rangle =0$, where $\{\cdot, \cdot\}$ denotes the anticommutator. A similar condition was also found in the earlier paper \cite{Johansen2004}, where it was expressed as 'vanishing current density'. However, we wish to obtain the strict equality and therefore we pose the more restrictive condition $\langle \phi\vert QP\phi\rangle =\frac{i}{2}$. Such a condition is satisfied for instance by a Gaussian $\phi (x) = \frac{1}{\sqrt{\Delta\sqrt{2\pi}}} e^{-\frac{x^2}{4\Delta^2}}$. Under these assumptions Eq. \eqref{limit} is clearly valid which shows that at least the real part of the (generalized) weak value is accessible via measurements. We summarize these considerations in the form of a proposition.

\begin{proposition}
Let $\h I^\lambda $ be the instrument defined by  the measurement scheme $\h M^\lambda= \langle \h K,  \phi,  \Phi^\lambda,  \sfp^Q, f^\lambda \rangle $ where $\phi\in \h D(QP) \cap \h D(PQ)$ is such that $\langle \phi \vert Q\phi\rangle=0$ and $\langle \phi \vert QP\phi\rangle = \frac{i}{2} $. Then 
\begin{equation}
\lim_{\lambda\rightarrow 0} \int x\,   \frac{\big\langle \varphi \big\vert \h I^\lambda (dx)^* (\F (Y)) \varphi \big\rangle}{\big\langle \varphi \big\vert \h I^\lambda (\R)^* (\F (Y)) \varphi \big\rangle}   =\mathrm{Re}\, \frac{\big\langle \varphi\big\vert \F(Y) \E[1]\varphi \big\rangle}{\big\langle\varphi\big\vert \F(Y)\varphi\big\rangle}
\end{equation}
for all $\varphi\in \widetilde{\h D} (x,\E)$.
\end{proposition}

In order to obtain the imaginary part we need another measurement setup. For that consider a measurement scheme $\h N^\lambda =\langle \h K,  \phi,  \Phi^\lambda,  \sfp^P, f^\lambda \rangle$, which differs from $\h M^\lambda$ by the pointer observable: instead of monitoring shifts in the probe's position we now observe the boosts the probe obtains. Note that at this point the initial probe state is also arbitrary. This change has a significant effect on the scheme. In particular, the measured observable becomes a trivial one, $X\mapsto \langle \phi\vert \sfp^P (\lambda X) \phi\rangle I$, meaning that no information is gained about the object system. However, this does not mean that the measurement does not affect the state of the system. Indeed, the instrument $\h J^\lambda$ associated with this scheme is nontrivial as can be seen from the dual form
$$
\h J^\lambda (X)^*(B) = \int_X \vert \hat{\phi} (x) \vert^2 \, V_\psi^* e^{i\lambda xA} (B\otimes I) e^{-i\lambda x A} V_\psi \, dx, 
$$
for all $B\in\lh$.

Now suppose that we again perform a sequential measurement where the second observable is $\F$. Then the calculations for the conditional average value and the limit of zero interaction are performed as before. In particular, by denoting $\Gamma^\lambda (X)  = \langle \varphi \vert \h J^\lambda (X)^* (\F(Y)) \varphi \rangle$ and by assuming that $\phi\in\h D(P)$ with $\langle \phi \vert P \phi\rangle =0 $  we find that
$$
\lim_{\lambda \rightarrow 0} \Gamma^\lambda[1] = 2\, \langle\phi \vert P^2\phi \rangle\cdot\textrm{Im}\, \langle \F(Y) \varphi \vert \E[1] \varphi \rangle
$$
so that by assuming that $\langle \phi \vert P^2\phi\rangle =\frac{1}{2}$ we get the desired result. For the Gaussians we have $\langle \phi \vert P^2\phi\rangle = \frac{1}{4\Delta^2}$ so that the above condition is satisfied with the choice $\Delta^2 =\frac{1}{2}$ for the variance. Once again we summarize this as a proposition.

\begin{proposition}
Let $\h J^\lambda $ be the instrument defined by  the measurement scheme $\h N^\lambda =\langle \h K,  \phi,  \Phi^\lambda,  \sfp^P, f^\lambda \rangle $ where $\phi\in \h D(P^2) $ is such that $\langle \phi \vert P\phi\rangle= 0$ and $\langle \phi \vert P^2\phi\rangle = \frac{1}{2} $. Then 
\begin{equation}
\lim_{\lambda\rightarrow 0} \int x\,   \frac{\big\langle \varphi \big\vert \h J^\lambda (dx)^* (\F (Y)) \varphi \big\rangle}{\big\langle \varphi \big\vert \h J^\lambda (\R)^* (\F (Y)) \varphi \big\rangle}   = \mathrm{Im}\, \frac{\big\langle \varphi\big\vert \F(Y) \E[1]\varphi \big\rangle}{\big\langle\varphi
\big\vert \F(Y)\varphi\big\rangle}
\end{equation}
for all $\varphi\in \widetilde{\h D} (x,\E)$.
\end{proposition}

We have shown  that the notion of the weak value of an observable as given in Definition \ref{Def} is operationally meaningful in the sense that there exist sequential measurement schemes which give the real and imaginary parts as conditional averages in the limit of zero interaction strength.

\section{State reconstruction methods}\label{reconstruction}

\subsection{Reconstructing the wavefunction via weak measurements}

In \cite{Lundeen2011} Lundeen {\em et al.} reported an experiment where they claim to have measured the pointwise values of the wavefunction of a photon using weak measurements. Leaving aside the  claims concerning the localization of a photon, the state determination method nevertheless deserves attention. To avoid the question of photon localization, we consider a spin-$\frac{1}{2}$ particle. In the original experiment the part of the inner degree of freedom was played by the polarization of the photon.

Consider the position of the particle in, say, the $z$-direction so that the spacial part of the Hilbert space can be taken to be $\hil =L^2 (\R)$. Now as a 'probe' we take the spin-degree of freedom, $\h K =\C^2$, with the initial state $\vert 0\rangle$. In order to measure the pointwise value of the wavefunction we divide the position space into disjoint intervals $(I_i)$ with the assumption that the intervals are of equal length and the center is labelled by $x_i$. For each interval we will then perform a weak measurement of the two-valued observable  $1\mapsto Q_i= \sfp^Q(I_i), 0\mapsto I-Q_i=\sfp^Q(\R\setminus I_i) $ thus scanning the whole position space. In order to accomplish the weak measurement, the position and spin of the particle are coupled via the unitary transformation
$$
\Phi^\alpha( \rho\otimes \sigma) = e^{-i\alpha Q_i\otimes \sigma_y}  \rho\otimes \sigma  e^{i\alpha Q_i\otimes \sigma_y}. 
$$
In particular, an initial vector state $\varphi \otimes \vert 0\rangle$ evolves into a superposition
\begin{equation}
\Psi^\alpha=Q_i\varphi \otimes \left(\cos\alpha\vert 0\rangle + \sin\alpha\vert 1\rangle\right) + (I-Q_i)\varphi \otimes \vert 0\rangle. 
\end{equation}
As the pointer observable we choose either $\sigma_x$ or $\sigma_y$ or, more precisely, their two-valued spectral measures $\sfp^{\sigma_j}$, $j=x,y$. We have thus arrived at the measurement schemes $\h M^i_j=\langle  \h K, \vert 0\rangle  , \Phi^\alpha  ,   \sfp^{\sigma_j}  \rangle$, $i\in\N$, $j=x,y$.

After the first measurement we measure the momentum $\sfp^P$ of the system and postselect the values which lie in the small interval $J_\epsilon =(-\frac{\epsilon}{2},\frac{\epsilon}{2})$. The conditional probabilities are then given by 
\begin{equation}
\{ \pm 1\}\mapsto\frac{\big\langle \Psi^\alpha\big\vert \sfp^P (J_\epsilon) \otimes \sfp^{\sigma_j} ( \{\pm 1\}) \Psi^\alpha\big\rangle}{\big\langle \Psi^\alpha\big\vert \sfp^P (J_\epsilon) \otimes I \Psi^\alpha\big\rangle}. 
\end{equation}
Now if we calculate the conditional average, and perform additional scaling by the factor $2\sin\alpha$, then in the weak limit $\alpha\rightarrow 0$ we arrive at the two values
\begin{eqnarray}
\xi_{i} &=& \textrm{Re} \langle \varphi\vert \sfp^P (J_\epsilon) Q_i\varphi\rangle \label{X}\\
\eta_{i} &=& \textrm{Im} \langle \varphi\vert \sfp^P (J_\epsilon)Q_i \varphi\rangle \label{Y}
\end{eqnarray}
where $\xi_{i}$ and $\eta_{i} $ refer to the measurements of $\sigma_x$ and $\sigma_y$, respectively, with  the fixed position interval $I_i$. The claim of  \cite{Lundeen2011} is that the quantities \eqref{X} and \eqref{Y} give the real and imaginary parts of the wavefunction at the point $x_i$, that is,
\begin{equation}\label{wavefunction}
\varphi(x_i) \simeq \textrm{constant}\cdot  \langle \varphi\vert \sfp^P (J_\epsilon) Q_i\varphi\rangle. 
\end{equation}
This is the more precise meaning of Eq. (2) and (3) in \cite{Lundeen2011}. It is now straightforward to show that for a sufficiently regular $\varphi$, such as a compactly supported $C^\infty$-function with $\hat{\varphi}(0)\neq 0$, Eq. \eqref{wavefunction} is true when the lengths of the intervals approach zero.

The proposed method has  obvious limitations as a method of state determination. Even after scanning through all the disjoint intervals $I_i$, the method can succeed only for those  state vectors $\varphi$ for which $\sfp^P (J_\epsilon)\varphi\neq 0$. Indeed, if the momentum of the system is localized outside the vicinity of the origin then $\sfp^P (J_\epsilon)\varphi  =0$ and no information can be obtained from the measurement. Thus, this method is far from being generally valid. Moreover, if it happens for instance that $\sfp^P (J_\epsilon)\varphi  =\varphi$, then it is known from the basic Fourier theory that all the component vectors $Q_i\varphi$, $i\in\N$, are nonzero and one has to scan through all the intervals $I_i$ in order to determine the state.

\subsection{Informationally complete sequential measurements}

As an alternative, and completely general method of state reconstruction we present the more common approximate sequential measurement of position and momentum where the weak limit is not taken. Consider again the standard measurement scheme $\h M^\lambda$ with the exception that we take the position $\sfp^Q$ to be the observable we wish to measure so that no dilation is needed and the transformation $\Phi^\lambda$ becomes the usual unitary transformation. The observable actually measured is thus a smeared position $\mu^\lambda*\sfp^Q$.

Suppose that after the measurement $\h M^\lambda$ a momentum measurement is performed. This results in a sequential joint observable $\M:\h B(\R^2)\rightarrow \lh$ which is in fact a {\em covariant phase space observable}. The structure and properties of such observables have been studied extensively and are well understood \cite{Holevo1979, Werner1984}. Indeed, the observable $\M$ is neatly presented as
\begin{equation}
\M (Z) = \frac{1}{2\pi} \int_Z W_{qp} \vert \phi^\lambda \rangle\langle \phi^\lambda \vert W_{qp}^*\, dqdp
\end{equation}
where $\phi^\lambda (x) =\sqrt{\lambda} \, \overline {\phi (-\lambda x)}$ and $W_{qp}=e^{i\frac{qp}{2}} e^{-iqP} e^{ipQ}$ are the Weyl operators. As proved in \cite{Ali1977}, if the initial probe state satisfies $\langle \phi^\lambda \vert W_{qp} \phi^\lambda\rangle\neq 0$ for almost all $(q,p)\in\R^2$, then the observable $\M$ is {\em informationally complete} \cite{Prugovecki1977}, meaning that the initial state of the system is uniquely determined by the measurement outcome statistics of this observable. This implies that by choosing the probe state to be for instance a Gaussian, it is possible to uniquely determine the state of the object system with a {\em single } measurement scheme. A well-known optical implementation of this measurement scheme is provided by an 8-port homodyne detector with a strong local oscillator. For a detailed analysis, see e.g. \cite{Kiukas2008}.

\section{Conclusions}\label{conclusion}
We have given a general definition of a weak value of a quantum observable and shown how it is related to sequential measurements of pairs of observables. In particular, we have shown that the real and imaginary parts of the weak value can be obtained as limits of conditional averages. We have also considered the determination of an unknown quantum state using both weak and approximate measurements. We have seen that the state reconstruction method of \cite{Lundeen2011} lacks the generality of the known methods of phase space measurements. 

\

\noindent
\textbf{Acknowledgment.} We are grateful to Dr. Mohamed Bourennane for drawing our attention to the paper by Lundeen {\em et al.} This work was supported by the Academy of Finland grant no. 138135. JS was supported by the Finnish Cultural Foundation.

\end{document}